\documentclass[twocolumn,secnumarabic,amssymb, nobibnotes, aps, prd]{revtex4-1}

\usepackage{graphicx}
\usepackage{amsmath}

\begin{document}

\title[Understanding magnetic focusing in graphene \textit{p-n} junctions through quantum modeling]
  {Understanding magnetic focusing in graphene \textit{p-n} junctions through quantum modeling}
  
\author{Samuel W LaGasse}%
\email[Samuel W LaGasse: ]{slagasse@sunypoly.edu}
\affiliation{Colleges of Nanoscale Science and Engineering, SUNY Polytechnic Institute, Albany, New York, 12203}
\author{Ji Ung Lee}%
\email[Ji Ung Lee: ]{jlee1@sunypoly.edu}
\affiliation{Colleges of Nanoscale Science and Engineering, SUNY Polytechnic Institute, Albany, New York, 12203}
\date{December 15, 2016}%

\keywords{Graphene, graphene \textit{p-n} junctions, magnetic focusing, quantum transport}

\begin{abstract}
We present a quantum model which provides enhanced understanding of recent transverse magnetic focusing experiments on graphene \textit{p-n} junctions. Spatially resolved flow maps of local particle current density show quantum interference and \textit{p-n} junction filtering effects which are crucial to explaining the device operation. The Landauer-B\"{u}ttiker formula is used alongside dephasing edge contacts to give exceptional agreement between simulated non-local resistance and the recent experiment by Chen \textit{et al} (\textit{Science}, 2016). The origin of positive and negative focusing resonances and off resonance characteristics are explained in terms of quantum transmission functions. Our model also captures subtle features from experiment, such as the previously unexplained $p$\textit{-}$p^-$ to $p$\textit{-}$p^+$ transition and the second \textit{p-n} focusing resonance. 
\end{abstract}

\maketitle

\section{Introduction}

%\section{Introduction}

Traditionally, transverse magnetic focusing (TMF) experiments have been restricted to unipolar conduction, in mediums such as metals \cite{Tsoi1978} and two-dimensional electron gasses (2DEG) \cite{VanHouten1989}. The discovery of graphene \cite{Novoselov2016}, in which electrons behave as massless Dirac fermions \cite{Novoselov2005}, has provided an exciting new platform for studying TMF. Graphene's gapless band structure, allowing ambipolar conduction, has enabled several recent TMF experiments. TMF in graphene has been studied as a function of carrier density \cite{Taychatanapat2013} and imaged with scanning gate microscopy \cite{Bhandari2016}. In addition, a large number of TMF peaks have been observed in graphene/hexagonal boron nitride superlattices \cite{Lee2016}. Recently, \textit{p-n} junctions in graphene have been used in TMF experiments to steer the focused beam \cite{Chen2016a}, opening the door to new electron optics. The \textit{p-n} junction is a fundamental device and has received a significant amount of attention from the graphene community. Graphene \textit{p-n} junctions have rich physical properties, exhibiting chiral tunneling \cite{Katsnelson2006,Young2008}, angle dependent transmission \cite{Cheianov2006,Sutar2012,Sajjad2012}, quantized conductance in high magnetic fields \cite{Abanin2007,Williams2007,NikolaiN.Klimov2015a}, and ballistic interference \cite{Rickhaus2013}. 

\begin{figure}
\includegraphics[width=0.925\columnwidth]{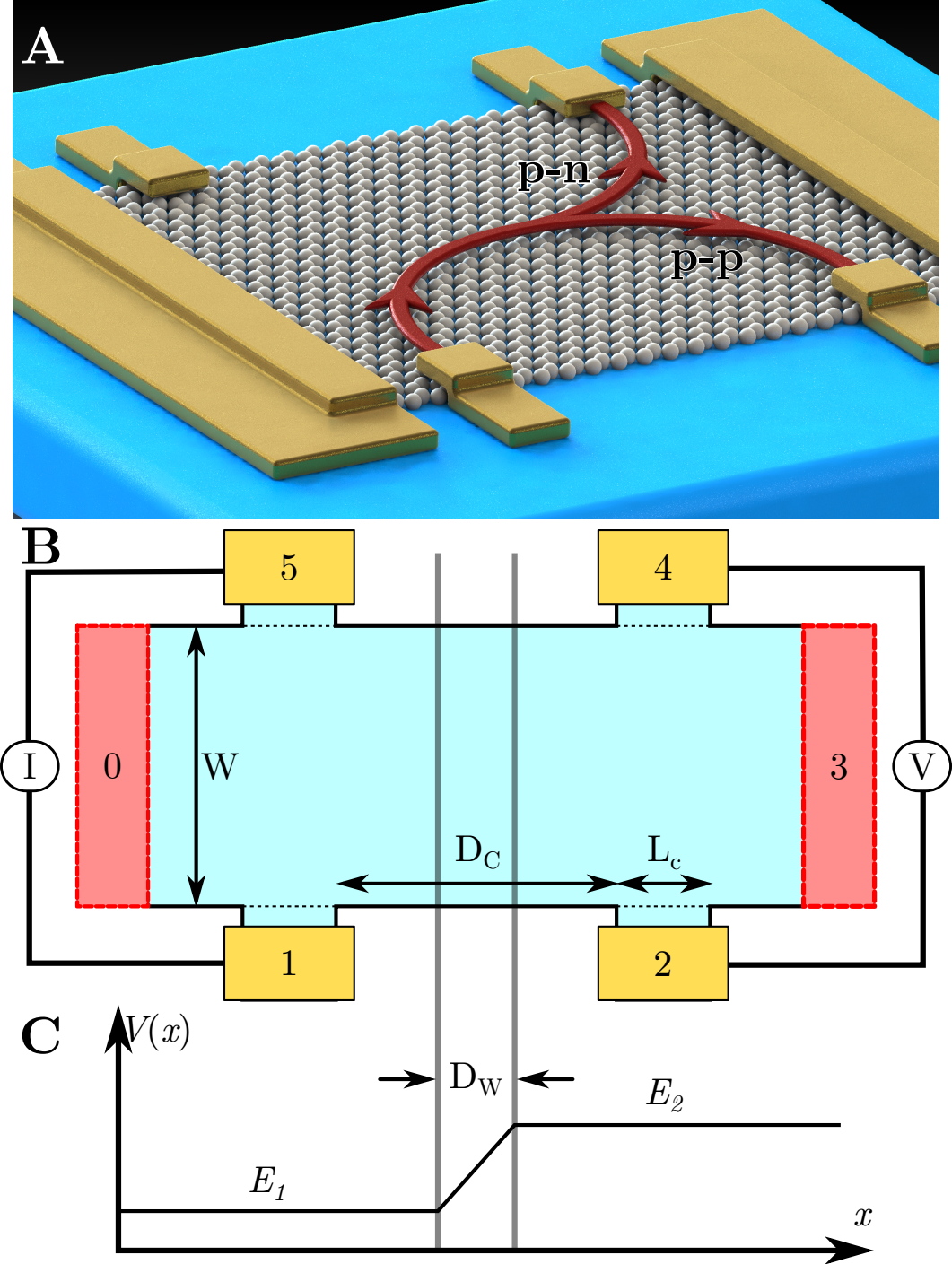}
	\caption{(a) Model of the graphene device depicting the first TMF resonance for \textit{p-p'} and \textit{p-n} junctions. (b) Schematic of device with four terminal measurement configuration. The device simulated has dimensions $D_C=W=200$ nm, $D_W=50$ nm, and $L_C=60$ nm. The red rectangles indicate dephasing contacts used in the simulation. (c) Real space energy band diagram of the device.}\label{fig:schematic}
\end{figure}
\begin{figure*}
\includegraphics[width=0.925\textwidth]{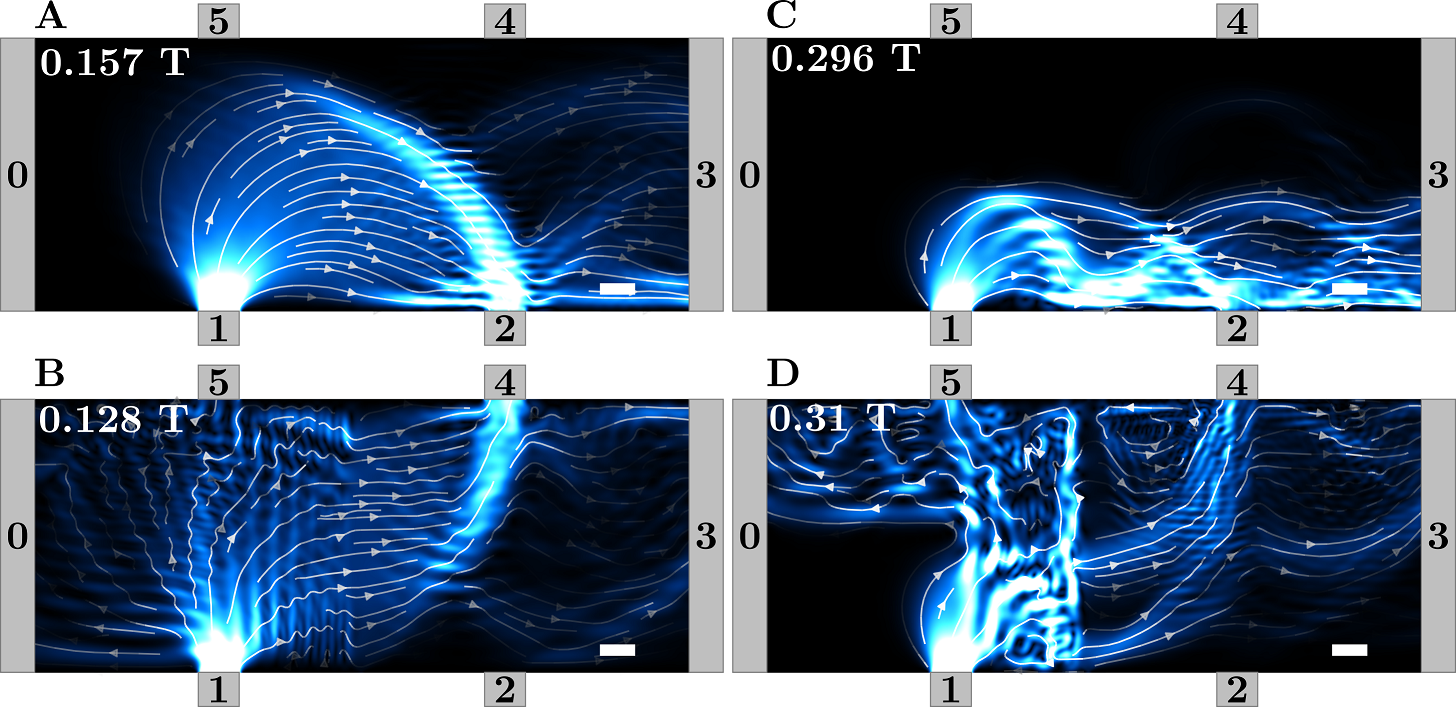}
	\caption{Maps of local particle current density (\ref{eq:CD}) for a \textit{p-p'} (\textit{p-n}) junction in (a) ((b)) the first TMF resonance and (c) ((d)) the second TMF resonance. The \textit{p-p'} junctions in (a) and (c) are configured as $E_1=-50$ meV and $E_2= -75$ meV. For the first \textit{p-n} TMF resonance in (b) the junction is configured as $E_1=-E_2=-50$ meV. The second \textit{p-n} TMF resonance in (d) is configured as $E_1=-50$ meV and $E_2=100$ meV. The scale bars are all 60 nm.}\label{fig:cd}
\end{figure*}

In this paper, we use quantum transport methods to model the graphene \textit{p-n} junction TMF experiment of Chen \textit{et al} \cite{Chen2016a}. Our calculations, implemented in the KWANT package \cite{Groth2014}, intrinsically capture quantum interference, tunneling, and angle dependent transmission \cite{Low2009b}, which enables us to explain the results of Chen \textit{et al} \cite{Chen2016a} in a completely quantum mechanical framework, without any fitting parameters. Previously, we have used the same basic model to understand quantum Hall measurements in graphene \textit{p-n} junctions \cite{LaGasse2016}. By including large dephasing edge contacts and performing multi-terminal Landauer-B\"{u}ttiker analysis\cite{Buttiker1988}, we are able to capture both the in-resonance and off-resonance characteristics of the device. We achieve exceptionally strong agreement between our simulation and experiment \cite{Chen2016a}, as shown in Fig. \ref{fig:map}. 

When a magnetic field is applied perpendicular to a graphene \textit{p-n} junction, electrons transporting across the junction will form snake states, arcing between the \textit{p} and \textit{n} sides of the junction \cite{Beenakker2008a}. In graphene, the arcs are characterized by the cyclotron radius, given by $r_c=\frac{\hbar\sqrt{\pi n}}{e|B|}$ with $\hbar$ the reduced Planck's constant, $n$ the carrier density, $e$ the electron charge, and $B$ the applied magnetic field. Snake states have been observed along graphene \textit{p-n} junctions in several experiments \cite{Williams2011,Rickhaus2015b,Taychatanapat2015}. Additionally, transport of electrons in snake states has been modeled using quantum mechanical\cite{Rickhaus2015b,Zarenia2013,Milovanovic2014a,Kolasinski2016} and semi-classical \cite{Carmier2010,Carmier2011,Patel2012,Davies2012,Milovanovic2013a,Milovanovic2014} methods. 

The TMF experiment performed on graphene \textit{p-n} junctions by Chen \textit{et al} \cite{Chen2016a} probes a special case of snake state transport, in a device similar to that depicted in Figures \ref{fig:schematic}a and \ref{fig:schematic}b. The device studied by Chen \textit{et al} \cite{Chen2016a} is special because the distance between contacts on each side of the junction, $D_C$, is approximately equal to the width of the device, $W$. When $2r_c \approx D_C$, the applied magnetic field focuses electrons directly between the contacts. In a unipolar system the carriers are directed back to the side from which they originate. Conversely, in a $p-n$ junction, the carriers will be steered towards the opposite side of the device. These two paths are depicted in Fig \ref{fig:schematic}a. 

\section{Transport Model}
In this paper, we study a tight-binding Hamiltonian describing low energy electrons in graphene, given by 

\begin{equation}\label{Hamiltonian}
\hat{H}= \sum_{i}^N\epsilon_i\hat{c}_i^\dag\hat{c}_i+\sum_{i,j}^Nt_{i,j}\hat{c}_i^\dag\hat{c}_j,
\end{equation}
where the second summation only takes place for atoms which are first nearest-neighbors. $\hat{c}_i^\dag / \hat{c}_j$ are Fermionic creation/annihilation operators, $\epsilon_i$ is the on-site energy at site \textit{i}, and $t_{i,j}$ is the hopping energy between sites \textit{i} and \textit{j}. The effect of an applied magnetic field is included using Peierl's substitution, $t_{i,j} = t_s\exp\left[i\frac{e}{\hbar}\int^{\textbf{r}_j}_{\textbf{r}_i} \textbf{A}\cdot\textbf{dr}\right]$, where we adopt a circular gauge for the vector potential $\textbf{A}$ \cite{Shevtsov2012}. We use a scaled tight-binding model\cite{Liu2015} where $a=s_{f}a_{0} \textrm{ and }  t_{s}=t_{0}/s_f$. The term $s_f=10$ scales the lattice constant, $a_0$, and the atomistic hopping energy , $t_0 \approx 2.7$ eV\cite{Reich2002}, to yield more efficient simulations. 

We simulate a six terminal Hall bar, as depicted in Fig \ref{fig:schematic}b, with four small contacts (labeled one, two, four, and five) and two large contacts (labeled zero and three). The spacing between the inner edges of the small contacts is set equal to the width of the Hall bar, $D_C=W$, which is the critical element of device design to observe the first \textit{p-n} focusing peak. To form \textit{p-n} junctions, the on-site energy on each side of the device may be tuned independently to $E_1$ and $E_2$. We set the on-site energy to change linearly between $E_1$ and $E_2$ over a junction width, $D_W$, as shown in the energy band diagram in Fig \ref{fig:schematic}c. 

The two large contacts, zero and three, are included as dephasing contacts. The voltages of these contacts are allowed to float in the simulation, accounting for any dephasing which occurs as the carrier wave skips along the left or right side of the device. This type of virtual dephasing contact has been used in quantum transport calculations in the past \cite{Stegmann2013}\cite{Stegmann2015} and is critical for tying our results to experiment.

Since most TMF measurements are performed at cryogenic temperatures under very small biases, we adopt a zero-bias, zero-temperature approximation. In this regime, we utilize the Landauer-B\"{u}ttiker equation\cite{Buttiker1988} to express the current in each lead \textit{p} \cite{Datta1995}, \begin{equation}\label{eq:LB}
I_p = \frac{2e^2}{h}\sum_q \left[ T_{qp}V_p-T_{pq}V_q\right], 
\end{equation} where the summation takes place over all leads in the system, including the dephasing contacts. For our simulation, (\ref{eq:LB}) generates a system of six linear equations with six unknowns. The term $T_{qp}$ is the quantum mechanical transmission function from lead \textit{p} to \textit{q}, defined as \begin{equation}\label{eq:transmission}
T_{qp}(E)=\sum_{n\in p, m\in q} \mid S_{nm}(E)\mid ^2,
\end{equation} where $S_{nm}$ is the scattering matrix element between the $n^{\textrm{th}}$ and $m^{\textrm{th}}$ mode in leads $p$ and $q$, respectively. The summation in (\ref{eq:transmission}) takes place over the available modes in each lead at energy $E$. 

To connect with the multi-terminal measurement of Chen \textit{et al} \cite{Chen2016a}, we simulate driving a current between contacts one and five and calculate the voltage acquired by contacts two and four. Practically, this requires setting $I_1=-I_5$, $I_0=I_2=I_3=I_4=0$, and choosing a contact to be grounded, in this case $V_1=0$. The non-local resistance for this configuration is defined as \begin{equation}\label{eq:nonlocal}
R_{15,24} = \frac{V_2-V_4}{I_1-I_5}.
\end{equation} The components of (\ref{eq:nonlocal}) are attained by solving the linear system, $\mathbf{I}=\frac{2e^2}{h}\mathbf{T}\mathbf{V}$, defined by (\ref{eq:LB}), where $\mathbf{I}$ and $\mathbf{V}$ are column vectors of lead currents and voltages, respectively, and $\mathbf{T}$ is a matrix of transmission functions. Making the substitutions above, (\ref{eq:nonlocal}) may be reduced to
$R_{15,24}=\frac{h}{2e^2}\frac{1}{2}\left(R_{45}-R_{25}\right)$. $R_{45}$ and $R_{25}$ are elements of the $R-$matrix, defined as $\mathbf{R}=\mathbf{T}^{-1}$, and are entirely comprised of transmission functions between different leads, thus, the problem is reduced to calculating the permutations of (\ref{eq:transmission}). 

To understand the terminal characteristics of our simulation, we generate spatially resolved particle current density maps using
\begin{equation}\label{eq:CD}
J_{\mathbf{r_i},\mathbf{r_j}}(E) = -2 \sum_{n\in p} \textrm{Im}\left[\psi_n(\mathbf{r_i},E)^\dagger \hat{H}_{i,j} \psi_n(\mathbf{r_j},E) \right]
\end{equation}
where $\mathbf{r_{i}}$ is the position of the $i^{\textrm{th}}$ lattice site, $\psi_n(\mathbf{r_i},E)$ is the wave function of the $n^{\textrm{th}}$ conducting mode in lead \textit{p}. The summation takes place over all conductive modes in lead \textit{p} available at energy $E$. However, separately resolving each mode is informative. 

\begin{figure*}
\includegraphics[width=0.95\textwidth]{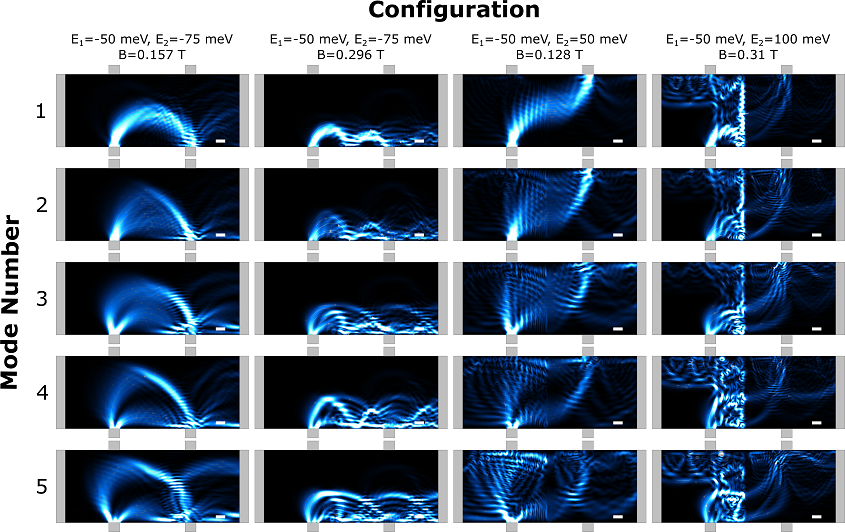}
	\caption{Table of mode resolved particle current density for each panel of Fig. \ref{fig:cd}. The modes of each column are summed to give the final result in Fig. \ref{fig:cd}. By looking at each mode individually, the interplay between the semi-classical and quantum mechanical nature of the system is visible.}\label{fig:mode_by_mode}
\end{figure*}

\begin{figure}
\includegraphics[width=0.95\columnwidth]{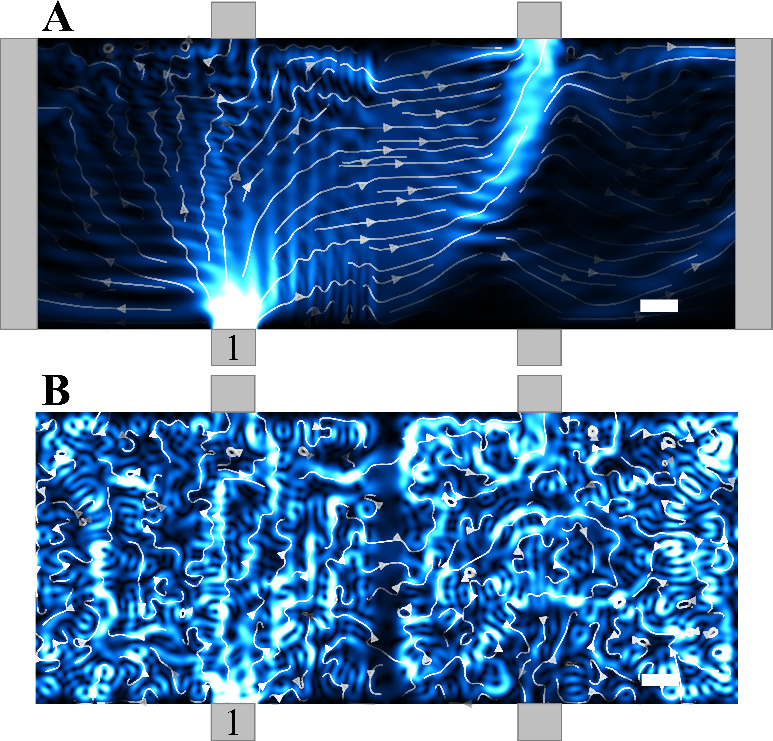}
	\caption{Comparison of particle current density for the device configured as in Fig. \ref{fig:cd}b both (a) with and (b) without dephasing edge contacts. When the dephasing edge contacts are removed, in (b), carrier density which is not focused into contact four will skip around the edge of the device until it exits out one of the small contacts. The carrier density, which is not dephased, will interfere with the incoming waves and destroy the resonance condition. This results in the extremely chaotic pattern seen in (b), with no observable focusing resonances.}\label{fig:dephase_comparison}
\end{figure}

\begin{figure}[ht]
\includegraphics[width=1.0\columnwidth]{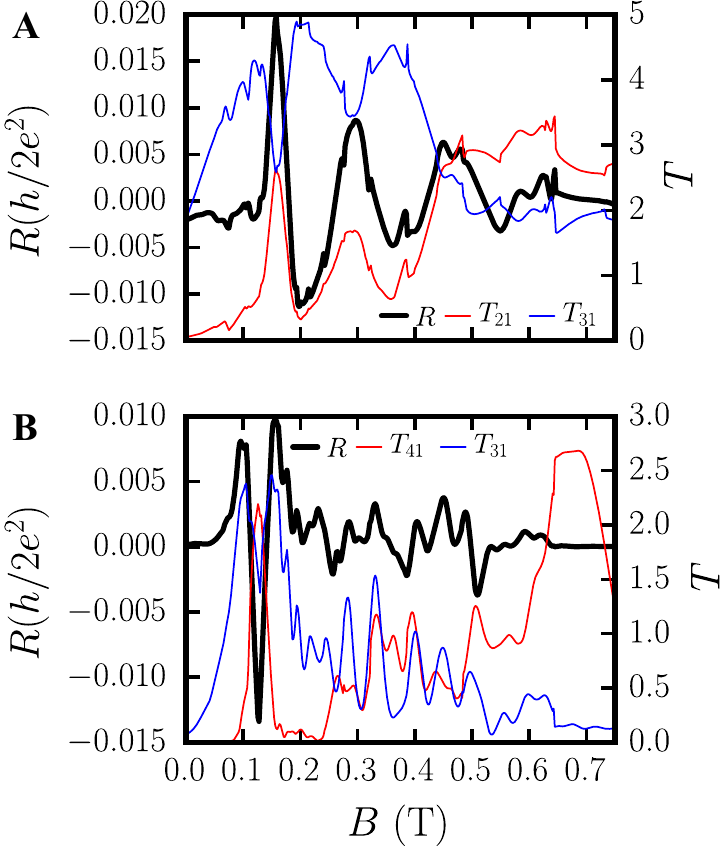}
	\caption{Non-local resistance as a function of magnetic field in an (a) \textit{p-p'} and (b) \textit{p-n} junction. The junctions are configured as in Fig \ref{fig:cd}. To compare to the carrier densities shown in Fig \ref{fig:map}, the left side of (a) is set to $-0.18\times10^{12} \textrm{ cm}^{-2}$ and the right side is set to $-0.41\times10^{12} \textrm{ cm}^{-2}$. The left side of (b) is configured the same as (a) but the right side is now set to $+0.18\times10^{12} \textrm{ cm}^{-2}$. Important transmission (\ref{eq:transmission}) functions are plotted for each configuration, as explained in the text.}\label{fig:linescan}
\end{figure}

\begin{figure}
\includegraphics[width=0.95\columnwidth]{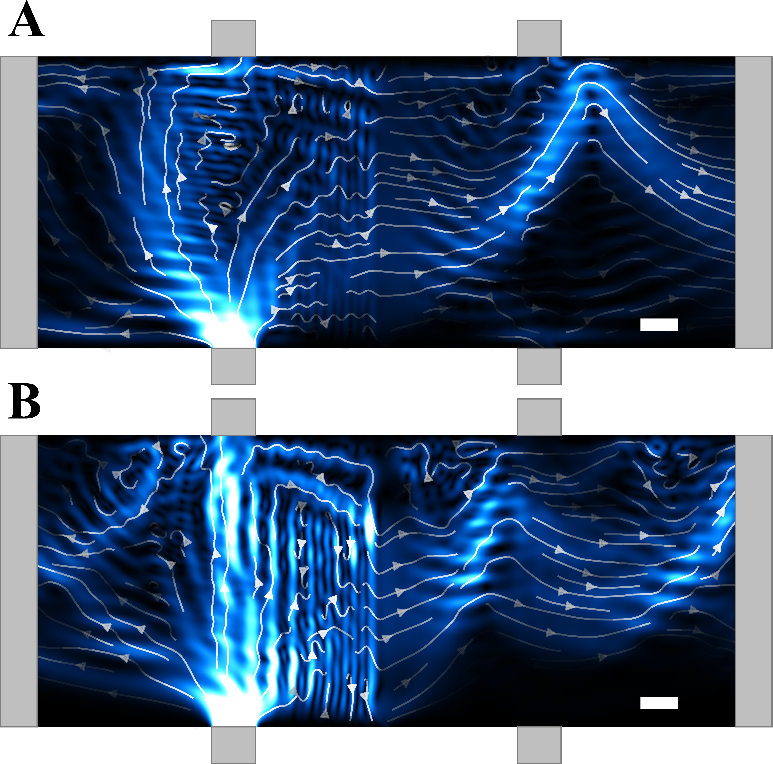}
	\caption{Off-resonance condition particle current density for the symmetric \textit{p-n} junction studied in Fig. \ref{fig:cd}b. The junction is configured as $E_1=-E_2=-50$ meV. In (a) $B=0.11$ T and in (b) $B=0.16$ T. When the magnetic field is not strong enough to focus the carriers into contact four, the transmitted wave collides with the top of the device, to the right of contact four, and skips into contact three. When the magnetic field is too strong, as in (b), the transmitted wave is focused to the left of contact four and again skips into contact three. }\label{fig:off_resonance_map}
\end{figure}

\section{Results and discussion}
%%%%%%%%%%%%%%%%%%%%%%%%%%%%%%%%%%%%%%%%%%%%%%%%%%%%%%%%%%%%%%%%
% Current density discussion
%%%%%%%%%%%%%%%%%%%%%%%%%%%%%%%%%%%%%%%%%%%%%%%%%%%%%%%%%%%%%%%%
In Fig. \ref{fig:cd} we plot vector flow maps of the local particle current density (\ref{eq:CD}) injected by contact one for \textit{p-p'} and \textit{p-n} junctions. When current is focused into contact two/four we observe positive/negative peaks in the non-local resistance, respectively. 

When the \textit{p-p'} junction is in the first TMF resonance, in Fig. \ref{fig:cd}a, carriers injected by contact one are focused directly into contact two. The carriers are injected and take on a broad spread of angles in the channel, but are primarily focused into a bright caustic which enters contact two. The junction redirects the carriers slightly, elongating the orbit. Due to the small size of the contacts, not all carriers which are injected by contact one are collected at contact two. Some hit the bottom edge of the device and skip into contact three, from which corresponding interference fringes may be seen, especially on the caustic. 

For the first resonance of the \textit{p-n} junction, in Fig. \ref{fig:cd}b, current injected from contact one is focused directly into contact four. The 50 nm junction width acts as a low pass filter, allowing only current flowing close to normal to the junction to transmit. On the left side of the junction, wave interference patterns indicate the current density reflected off the junction, which then exits out the contacts on the left side of the device. The transmitted current predominately focuses into a caustic which enters contact four. 

At low magnetic fields, a significant portion of the current injected from contact one hits the top edge of the device before crossing the junction, as seen for \textit{p-n} junction in Fig. \ref{fig:cd}b. This is a consequence of the device geometry studied by Chen \textit{et al} \cite{Chen2016a} and increasing the device width to avoid hitting the top edge prohibits one from probing the first \textit{p-n} TMF resonance. Interestingly, a component of the current hitting the top edge is redirected and transmits across the junction. This subtle detail, captured by our model, contributes to the device's terminal characteristics and is important in many of the different junction configurations. 

When the magnetic field is increased to the second TMF resonance, in Fig. \ref{fig:cd}c and d, the current density will skip along the edge of the junction (\textit{p-n} case) or the edge of the device (\textit{p-p'} case). \textit{p-n} junctions do not exhibit the second resonance until the \textit{n-}doping is stronger than the \textit{p-}doping, thus we configure the junction in Fig. \ref{fig:cd}d as $E_1=-50$ meV and $E_2=100$ meV. In the \textit{p-n} configuration, on the \textit{p}-side, the current forms a circular orbit which reflects near the bottom of the junction and again almost half way up. At each of these points there is a significant portion of current which is incoming normal to the junction and transmits to the other side, focusing on contact 4. Due to the filtering effect of a smooth \textit{p-n} junction, the second TMF resonance is significantly weaker. 

%%%%%%%%%%%%%%%%%%%%%%%%%%%%%%%%%%%%%%%%%%%%%%%%%%%%%%%%%%%%%%%%
% Mode by mode discusssion
%%%%%%%%%%%%%%%%%%%%%%%%%%%%%%%%%%%%%%%%%%%%%%%%%%%%%%%%%%%%%%%%

To further understand the local particle current density of the devices in Fig. \ref{fig:cd}, in Fig. \ref{fig:mode_by_mode} we resolve the characteristic by each propagating mode. By resolving each mode which contributes to the results in Fig. \ref{fig:cd}, we observe a combination of features reminiscent of semi-classical skipping orbits and quantum mechanical interference patterns.

The lowest mode is injected straight into the device, perpendicular to the semi-infinite contact. In the first resonance of the \textit{p-p'} and \textit{p-n} junction, shown in columns one and three of Fig. \ref{fig:mode_by_mode}, respectively, the lowest mode is bent so that the wave is propagating approximately normal to the junction when it crosses it. Thus, the lowest mode is nearly perfectly transmitted, with very few reflections (indicated by interference fringes) visible. 

Higher modes are injected into the device with non-zero angles and arrive at the junction traveling at oblique angles. For the first resonance of the \textit{p-p'}, the beam is noticeably refracted as it crosses the junction. In the \textit{p-n} junction, the higher order modes have significant components which are reflected off the junction, due to the angle dependent transmission across the junction.

For the second resonance of the \textit{p-p'} and \textit{p-n'} configurations, the local particle current density patterns in Fig. \ref{fig:cd} are more complex than the first resonance. By resolving each mode, we are able to develop a better picture of the important transport mechanisms. The higher order modes for the \textit{p-p'} junction have a component which transports nearly parallel to the lower edge of the device. This is particularly evident in the fourth and fifth modes. Most of the carriers which transport in this manner will miss contact two and transmit out contact three, resulting in a weaker signal for the second focusing resonance. 

The second focusing resonance of the \textit{p-n'} displays the most complex characteristics of the device, with predominant quantum characteristics not present in the other configurations. At the higher magnetic field, the first and second modes appear to begin to form Landau levels when they collide with the junction, similar to what we have studied in our previous work \cite{LaGasse2016}. The higher order modes, however, instead show a more complex, swirling pattern. The carriers transport in skipping orbits which partially reflect of the junction, interfering with themselves. A portion of each orbit transmits across the junction, contributing to the second \textit{p-n'} resonance.  

%%%%%%%%%%%%%%%%%%%%%%%%%%%%%%%%%%%%%%%%%%%%%%%%%%%%%%%%%%%%%%%%
% Dephasing edge contact discussion
%%%%%%%%%%%%%%%%%%%%%%%%%%%%%%%%%%%%%%%%%%%%%%%%%%%%%%%%%%%%%%%%
As mentioned previously, the dephasing edge contacts (labeled contact zero and three) are critical to attaining the results presented in this paper. To demonstrate this importance, in Fig. \ref{fig:dephase_comparison} we plot the local particle current density for the device configured as in Fig. \ref{fig:cd}b both with and without the dephasing contacts. When the dephasing contacts are removed, in Fig. \ref{fig:dephase_comparison}b, the portions of the wave which normally exit contacts zero and three, instead scatters around the edge of the device. The wave will continue to scatter around the device, interfering with itself, until exiting out one of the small contacts. This process occurs until the device reaches steady state, resulting in the extremely chaotic pattern shown and the destruction of any resonance characteristics. 
%%%%%%%%%%%%%%%%%%%%%%%%%%%%%%%%%%%%%%%%%%%%%%%%%%%%%%%%%%%%%%%%
% Linescan discussion
%%%%%%%%%%%%%%%%%%%%%%%%%%%%%%%%%%%%%%%%%%%%%%%%%%%%%%%%%%%%%%%%

Fig. \ref{fig:linescan} shows the non-local resistance (\ref{eq:nonlocal}) and selected transmission coefficients (\ref{eq:transmission}) as a function of applied magnetic field for an asymmetric \textit{p-p'} junction and a symmetric \textit{p-n} junction. The two junction configurations are doped the same as in Fig. \ref{fig:cd}a and b, respectively. 

It is non-trivial to extract specific terms from (\ref{eq:nonlocal}), in terms of transmission functions, which result in the final form of the non-local resistance. The final magnitude and shape of the curve consists of permutations of transmission functions between every contact combined together. However, we are able to target specific transmission functions which are important in understanding the problem. 

In the unipolar \textit{p-p'} configuration, we observe three well defined TMF resonances. We are able to match the first two TMF resonances to a peak in the transmission from contact one into two, $T_{21}$. When the junction is switched to the \textit{p-n} configuration, when in resonance, carriers are now focused from contact one into contact four. This results in a negative peak in resistance at $B=0.128$ T, shown in Fig. \ref{fig:linescan}b. The important transmission function for understanding the resonance condition of the \textit{p-n} junction is $T_{41}$, which is peaked while the device is in resonance.

Each subsequent TMF resonance of the unipolar junction configuration decreases in magnitude. For higher order TMF resonances, an increase in $T_{31}$ indicates that the focusing effect is diminished. This is due to interference caused by the increased number of scattering events off the edges of the device. 

When either configuration of junction is not in resonance, there is an increase in the transmission from contact one into contact three, $T_{31}$. In the off-resonance state of the \textit{p-p'} junction, carriers which are not focused from contact one into two will hit the bottom edge of the device and skip into contact three. To maintain current conservation, the carriers will be re-injected by the floating contact three and the magnetic field will direct the carriers towards contact four, resulting in the negative off-resonance resistance in Fig. \ref{fig:linescan}a. Conversely, in the \textit{p-n} junction, carriers which miss contact four will skip along the top edge of the device. Again, they will be dephased by contact three, except this time the re-injected carriers will be directed towards contact two, which results in the positive off-resonance resistance in Fig. \ref{fig:linescan}b. In Fig. \ref{fig:off_resonance_map} we illustrate the off-resonance particle current density for a symmetric \textit{p-n} junction.

\begin{figure}[ht!]
\includegraphics[width=0.95\columnwidth]{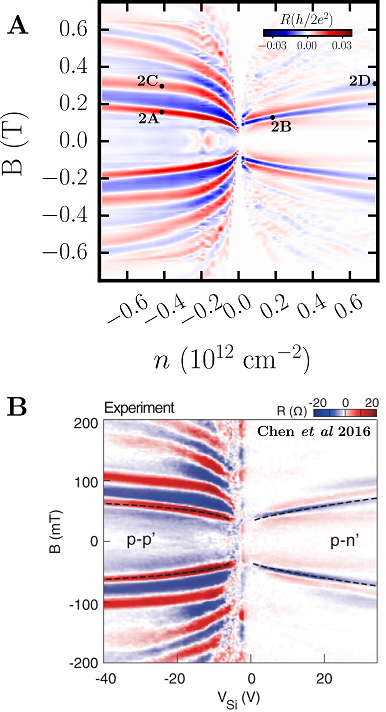}
	\caption{(a) Non-local resistance map for a fixed \textit{p}-type doping as a function of the carrier density of the right side of the junction and magnetic field. Scatter points mark configurations where we have demonstrated spatially resolved particle current density in Fig. \ref{fig:cd}. Our results show exceptional agreement with (b) experimental measurements of Chen \textit{et al} \cite{Chen2016a}, reproduced with copyright permission.}\label{fig:map}
\end{figure}

The junction filtering effect, seen in Fig. \ref{fig:cd}, results in significantly weaker and fewer TMF resonances when the device is in the \textit{p-n} configuration. For the symmetric \textit{p-n} junction, in Fig. \ref{fig:linescan}, only a single well defined resonance is observed. At higher magnetic fields the beam of carriers skips along the edge of the junction; each time the beam hits the junction only a very small amount will leak through. Since, in our model, no dephasing happens along the junction, the reflected  wave of carriers will interfere with itself, further disrupting any resonance from setting up. 

In addition, the non-local resistance tends towards zero for each configuration at around $B=0.65$ T. This is due to the carriers being forced into edge states as the device enters the quantum Hall regime. This effect is also the reason why we do not see a well defined peak in $T_{21}$ for the third TMF resonance of the \textit{p-p'} junction configuration. Capturing this feature highlights the power of quantum transport modeling, where our simulations smoothly transition between carriers occupying semi-classical skipping orbits and edge states.  

%%%%%%%%%%%%%%%%%%%%%%%%%%%%%%%%%%%%%%%%%%%%%%%%%%%%%%%%%%%%%%%%
% Map discussion
%%%%%%%%%%%%%%%%%%%%%%%%%%%%%%%%%%%%%%%%%%%%%%%%%%%%%%%%%%%%%%%%
Finally, in Fig. \ref{fig:map}, we compare our model with the recent experimental data of Chen \textit{et al} \cite{Chen2016a}, reproduced with copyright permission. In Fig. \ref{fig:map}a, we fix the doping of the left side of the junction to $E_1=-50$ meV (p-type) and vary the doping of the right side of the junction and applied magnetic field simultaneously. For each configuration we calculate the non-local resistance (\ref{eq:nonlocal}) as before. We report the doping of the right side in terms of carrier density $n$, which has a similar functional form to the gate voltage applied in experiment. 

Our simulation results show a striking similarity to the experimental data, capturing all of the major features. These include the four unipolar junction TMF resonances, the first ambipolar TMF resonance, and the negative/positive peaks in resistance when the unipolar/ambipolar configurations are not in resonance, respectively. 

We also are able to explain a number of subtle features seen experimentally which are due to the transitions between different types of junctions. The second negative peak in the unipolar junction configuration begins to disappear as the right side of the device is more strongly doped \textit{p}-type. This transition occurs when the doping of the right side of the junction exceeds the doping of the left side.  

The second ambipolar junction TMF resonance in Fig. \ref{fig:map} is extremely weak until the \textit{n}-type doping of the right side of the junction exceeds the fixed \textit{p}-type doping. This effect is enabled by the increased number of modes available to conduct on the right side of the junction as the doping is increased. The filtering effect due to the large junction width present in our model and in experiment \cite{Chen2016a} prohibits the traditional picture of the carrier density snaking across the junction several times in the second TMF resonance. Instead, the resonance has the characteristic of the flow map shown in Fig. \ref{fig:cd}d.

In our simulation a larger magnetic field must be used, since our simulated Hall bar is about a factor of ten smaller than the experimental device. The difference in device size and contact dimensions also accounts for the difference in magnitude of our simulated resistance. Using larger contacts will result in smaller values of resistance. However, the concepts we have discussed may still be applied to understand the experimental measurements of Chen \textit{et al} \cite{Chen2016a}.

\section{Conclusion}

In conclusion, we demonstrate a quantum transport model  for a TMF experiment on graphene \textit{p-n} by Chen \textit{et al} \cite{Chen2016a}.  Spatially resolved particle current density flow maps reveal the behavior of carriers in the first and second resonances of \textit{p-p'} and \textit{p-n} junctions. Our results demonstrate the importance of wave interference and junction filtering effect for understanding TMF experiments. A combination of dephasing edge contacts and use of the Landauer-B\"{u}ttiker formula  supplementing the standard tight-binding model yield extremely close agreement with experiment. Our non-local resistance simulations show well defined positive and negative peaks, which are due to enhanced transmission into contacts three or four, respectively. Many of the features seen by Chen \textit{et al} \cite{Chen2016a} have been explained, including the transition into the quantum Hall regime for high magnetic fields and the transitions between different \textit{p-p'} and \textit{p-n} doping regimes.

\begin{acknowledgments}
The authors acknolwedge financial support provided by the U.S. Naval Research Laboratory (Grant Number: N00173-14-1-G017). 

\end{acknowledgments}


\begin{thebibliography}{40}%
\makeatletter
\providecommand \@ifxundefined [1]{%
 \@ifx{#1\undefined}
}%
\providecommand \@ifnum [1]{%
 \ifnum #1\expandafter \@firstoftwo
 \else \expandafter \@secondoftwo
 \fi
}%
\providecommand \@ifx [1]{%
 \ifx #1\expandafter \@firstoftwo
 \else \expandafter \@secondoftwo
 \fi
}%
\providecommand \natexlab [1]{#1}%
\providecommand \enquote  [1]{``#1''}%
\providecommand \bibnamefont  [1]{#1}%
\providecommand \bibfnamefont [1]{#1}%
\providecommand \citenamefont [1]{#1}%
\providecommand \href@noop [0]{\@secondoftwo}%
\providecommand \href [0]{\begingroup \@sanitize@url \@href}%
\providecommand \@href[1]{\@@startlink{#1}\@@href}%
\providecommand \@@href[1]{\endgroup#1\@@endlink}%
\providecommand \@sanitize@url [0]{\catcode `\\12\catcode `\$12\catcode
  `\&12\catcode `\#12\catcode `\^12\catcode `\_12\catcode `\%12\relax}%
\providecommand \@@startlink[1]{}%
\providecommand \@@endlink[0]{}%
\providecommand \url  [0]{\begingroup\@sanitize@url \@url }%
\providecommand \@url [1]{\endgroup\@href {#1}{\urlprefix }}%
\providecommand \urlprefix  [0]{URL }%
\providecommand \Eprint [0]{\href }%
\providecommand \doibase [0]{http://dx.doi.org/}%
\providecommand \selectlanguage [0]{\@gobble}%
\providecommand \bibinfo  [0]{\@secondoftwo}%
\providecommand \bibfield  [0]{\@secondoftwo}%
\providecommand \translation [1]{[#1]}%
\providecommand \BibitemOpen [0]{}%
\providecommand \bibitemStop [0]{}%
\providecommand \bibitemNoStop [0]{.\EOS\space}%
\providecommand \EOS [0]{\spacefactor3000\relax}%
\providecommand \BibitemShut  [1]{\csname bibitem#1\endcsname}%
\let\auto@bib@innerbib\@empty
%</preamble>
\bibitem [{\citenamefont {Tsoi}\ and\ \citenamefont {Tsoi}(1978)}]{Tsoi1978}%
  \BibitemOpen
  \bibfield  {author} {\bibinfo {author} {\bibfnamefont {V.~S.}\ \bibnamefont
  {Tsoi}}\ and\ \bibinfo {author} {\bibfnamefont {N.}~\bibnamefont {Tsoi}},\
  }\href@noop {} {\bibfield  {journal} {\bibinfo  {journal} {Jetp}\ }\textbf
  {\bibinfo {volume} {1962}},\ \bibinfo {pages} {150} (\bibinfo {year}
  {1978})}\BibitemShut {NoStop}%
\bibitem [{\citenamefont {van Houten}\ \emph {et~al.}(1989)\citenamefont {van
  Houten}, \citenamefont {Beenakker}, \citenamefont {Williamson}, \citenamefont
  {Broekaart}, \citenamefont {van Loosdrecht}, \citenamefont {van Wees},
  \citenamefont {Mooij}, \citenamefont {Foxon},\ and\ \citenamefont
  {Harris}}]{VanHouten1989}%
  \BibitemOpen
  \bibfield  {author} {\bibinfo {author} {\bibfnamefont {H.}~\bibnamefont {van
  Houten}}, \bibinfo {author} {\bibfnamefont {C.~W.~J.}\ \bibnamefont
  {Beenakker}}, \bibinfo {author} {\bibfnamefont {J.~G.}\ \bibnamefont
  {Williamson}}, \bibinfo {author} {\bibfnamefont {M.~E.~I.}\ \bibnamefont
  {Broekaart}}, \bibinfo {author} {\bibfnamefont {P.~H.~M.}\ \bibnamefont {van
  Loosdrecht}}, \bibinfo {author} {\bibfnamefont {B.~J.}\ \bibnamefont {van
  Wees}}, \bibinfo {author} {\bibfnamefont {J.~E.}\ \bibnamefont {Mooij}},
  \bibinfo {author} {\bibfnamefont {C.~T.}\ \bibnamefont {Foxon}}, \ and\
  \bibinfo {author} {\bibfnamefont {J.~J.}\ \bibnamefont {Harris}},\ }\href
  {\doibase 10.1103/PhysRevB.39.8556} {\bibfield  {journal} {\bibinfo
  {journal} {Physical Review B}\ }\textbf {\bibinfo {volume} {39}},\ \bibinfo
  {pages} {8556} (\bibinfo {year} {1989})}\BibitemShut {NoStop}%
\bibitem [{\citenamefont {Novoselov}\ \emph {et~al.}(2004)\citenamefont
  {Novoselov}, \citenamefont {Geim}, \citenamefont {Morozov}, \citenamefont
  {Jiang}, \citenamefont {Zhang}, \citenamefont {Dubonos}, \citenamefont
  {Grigorieva},\ and\ \citenamefont {Firsov}}]{Novoselov2016}%
  \BibitemOpen
  \bibfield  {author} {\bibinfo {author} {\bibfnamefont {K.~S.}\ \bibnamefont
  {Novoselov}}, \bibinfo {author} {\bibfnamefont {A.~K.}\ \bibnamefont {Geim}},
  \bibinfo {author} {\bibfnamefont {S.~V.}\ \bibnamefont {Morozov}}, \bibinfo
  {author} {\bibfnamefont {D.}~\bibnamefont {Jiang}}, \bibinfo {author}
  {\bibfnamefont {Y.}~\bibnamefont {Zhang}}, \bibinfo {author} {\bibfnamefont
  {S.~V.}\ \bibnamefont {Dubonos}}, \bibinfo {author} {\bibfnamefont {I.~V.}\
  \bibnamefont {Grigorieva}}, \ and\ \bibinfo {author} {\bibfnamefont {A.~A.}\
  \bibnamefont {Firsov}},\ }\href {\doibase 10.1126/science.1102896} {\bibfield
   {journal} {\bibinfo  {journal} {Science}\ }\textbf {\bibinfo {volume}
  {306}},\ \bibinfo {pages} {666} (\bibinfo {year} {2004})}\BibitemShut
  {NoStop}%
\bibitem [{\citenamefont {Novoselov}\ \emph {et~al.}(2005)\citenamefont
  {Novoselov}, \citenamefont {Geim}, \citenamefont {Morozov}, \citenamefont
  {Jiang}, \citenamefont {Katsnelson}, \citenamefont {Grigorieva},
  \citenamefont {Dubonos},\ and\ \citenamefont {Firsov}}]{Novoselov2005}%
  \BibitemOpen
  \bibfield  {author} {\bibinfo {author} {\bibfnamefont {K.~S.}\ \bibnamefont
  {Novoselov}}, \bibinfo {author} {\bibfnamefont {A.~K.}\ \bibnamefont {Geim}},
  \bibinfo {author} {\bibfnamefont {S.~V.}\ \bibnamefont {Morozov}}, \bibinfo
  {author} {\bibfnamefont {D.}~\bibnamefont {Jiang}}, \bibinfo {author}
  {\bibfnamefont {M.~I.}\ \bibnamefont {Katsnelson}}, \bibinfo {author}
  {\bibfnamefont {I.~V.}\ \bibnamefont {Grigorieva}}, \bibinfo {author}
  {\bibfnamefont {S.~V.}\ \bibnamefont {Dubonos}}, \ and\ \bibinfo {author}
  {\bibfnamefont {A.~A.}\ \bibnamefont {Firsov}},\ }\href {\doibase
  10.1038/nature04233} {\bibfield  {journal} {\bibinfo  {journal} {Nature}\
  }\textbf {\bibinfo {volume} {438}},\ \bibinfo {pages} {197} (\bibinfo {year}
  {2005})}\BibitemShut {NoStop}%
\bibitem [{\citenamefont {Taychatanapat}\ \emph {et~al.}(2013)\citenamefont
  {Taychatanapat}, \citenamefont {Watanabe}, \citenamefont {Taniguchi},\ and\
  \citenamefont {Jarillo-Herrero}}]{Taychatanapat2013}%
  \BibitemOpen
  \bibfield  {author} {\bibinfo {author} {\bibfnamefont {T.}~\bibnamefont
  {Taychatanapat}}, \bibinfo {author} {\bibfnamefont {K.}~\bibnamefont
  {Watanabe}}, \bibinfo {author} {\bibfnamefont {T.}~\bibnamefont {Taniguchi}},
  \ and\ \bibinfo {author} {\bibfnamefont {P.}~\bibnamefont
  {Jarillo-Herrero}},\ }\href {\doibase 10.1038/nphys2549} {\bibfield
  {journal} {\bibinfo  {journal} {Nature Physics}\ }\textbf {\bibinfo {volume}
  {9}},\ \bibinfo {pages} {225} (\bibinfo {year} {2013})}\BibitemShut {NoStop}%
\bibitem [{\citenamefont {Bhandari}\ \emph {et~al.}(2016)\citenamefont
  {Bhandari}, \citenamefont {Lee}, \citenamefont {Klales}, \citenamefont
  {Watanabe}, \citenamefont {Taniguchi}, \citenamefont {Heller}, \citenamefont
  {Kim},\ and\ \citenamefont {Westervelt}}]{Bhandari2016}%
  \BibitemOpen
  \bibfield  {author} {\bibinfo {author} {\bibfnamefont {S.}~\bibnamefont
  {Bhandari}}, \bibinfo {author} {\bibfnamefont {G.-H.}\ \bibnamefont {Lee}},
  \bibinfo {author} {\bibfnamefont {A.}~\bibnamefont {Klales}}, \bibinfo
  {author} {\bibfnamefont {K.}~\bibnamefont {Watanabe}}, \bibinfo {author}
  {\bibfnamefont {T.}~\bibnamefont {Taniguchi}}, \bibinfo {author}
  {\bibfnamefont {E.}~\bibnamefont {Heller}}, \bibinfo {author} {\bibfnamefont
  {P.}~\bibnamefont {Kim}}, \ and\ \bibinfo {author} {\bibfnamefont {R.~M.}\
  \bibnamefont {Westervelt}},\ }\href {\doibase 10.1021/acs.nanolett.5b04609}
  {\bibfield  {journal} {\bibinfo  {journal} {Nano Letters}\ }\textbf {\bibinfo
  {volume} {16}},\ \bibinfo {pages} {1690} (\bibinfo {year}
  {2016})}\BibitemShut {NoStop}%
\bibitem [{\citenamefont {Lee}\ \emph {et~al.}(2016)\citenamefont {Lee},
  \citenamefont {Wallbank}, \citenamefont {Gallagher}, \citenamefont
  {Watanabe}, \citenamefont {Taniguchi}, \citenamefont
  {Fal{\textquoteright}ko},\ and\ \citenamefont {Goldhaber-Gordon}}]{Lee2016}%
  \BibitemOpen
  \bibfield  {author} {\bibinfo {author} {\bibfnamefont {M.}~\bibnamefont
  {Lee}}, \bibinfo {author} {\bibfnamefont {J.~R.}\ \bibnamefont {Wallbank}},
  \bibinfo {author} {\bibfnamefont {P.}~\bibnamefont {Gallagher}}, \bibinfo
  {author} {\bibfnamefont {K.}~\bibnamefont {Watanabe}}, \bibinfo {author}
  {\bibfnamefont {T.}~\bibnamefont {Taniguchi}}, \bibinfo {author}
  {\bibfnamefont {V.~I.}\ \bibnamefont {Fal{\textquoteright}ko}}, \ and\
  \bibinfo {author} {\bibfnamefont {D.}~\bibnamefont {Goldhaber-Gordon}},\
  }\href {\doibase 10.1126/science.aaf1095} {\bibfield  {journal} {\bibinfo
  {journal} {Science}\ }\textbf {\bibinfo {volume} {353}},\ \bibinfo {pages}
  {1526} (\bibinfo {year} {2016})}\BibitemShut {NoStop}%
\bibitem [{\citenamefont {Chen}\ \emph {et~al.}(2016)\citenamefont {Chen},
  \citenamefont {Han}, \citenamefont {Elahi}, \citenamefont {Habib},
  \citenamefont {Wang}, \citenamefont {Wen}, \citenamefont {Gao}, \citenamefont
  {Taniguchi}, \citenamefont {Watanabe}, \citenamefont {Hone}, \citenamefont
  {Ghosh},\ and\ \citenamefont {Dean}}]{Chen2016a}%
  \BibitemOpen
  \bibfield  {author} {\bibinfo {author} {\bibfnamefont {S.}~\bibnamefont
  {Chen}}, \bibinfo {author} {\bibfnamefont {Z.}~\bibnamefont {Han}}, \bibinfo
  {author} {\bibfnamefont {M.~M.}\ \bibnamefont {Elahi}}, \bibinfo {author}
  {\bibfnamefont {K.~M.~M.}\ \bibnamefont {Habib}}, \bibinfo {author}
  {\bibfnamefont {L.}~\bibnamefont {Wang}}, \bibinfo {author} {\bibfnamefont
  {B.}~\bibnamefont {Wen}}, \bibinfo {author} {\bibfnamefont {Y.}~\bibnamefont
  {Gao}}, \bibinfo {author} {\bibfnamefont {T.}~\bibnamefont {Taniguchi}},
  \bibinfo {author} {\bibfnamefont {K.}~\bibnamefont {Watanabe}}, \bibinfo
  {author} {\bibfnamefont {J.}~\bibnamefont {Hone}}, \bibinfo {author}
  {\bibfnamefont {A.~W.}\ \bibnamefont {Ghosh}}, \ and\ \bibinfo {author}
  {\bibfnamefont {C.~R.}\ \bibnamefont {Dean}},\ }\href {\doibase
  10.1126/science.aaf5481} {\bibfield  {journal} {\bibinfo  {journal}
  {Science}\ }\textbf {\bibinfo {volume} {353}},\ \bibinfo {pages} {1522}
  (\bibinfo {year} {2016})}\BibitemShut {NoStop}%
\bibitem [{\citenamefont {Katsnelson}\ \emph {et~al.}(2006)\citenamefont
  {Katsnelson}, \citenamefont {Novoselov},\ and\ \citenamefont
  {Geim}}]{Katsnelson2006}%
  \BibitemOpen
  \bibfield  {author} {\bibinfo {author} {\bibfnamefont {M.~I.}\ \bibnamefont
  {Katsnelson}}, \bibinfo {author} {\bibfnamefont {K.~S.}\ \bibnamefont
  {Novoselov}}, \ and\ \bibinfo {author} {\bibfnamefont {A.~K.}\ \bibnamefont
  {Geim}},\ }\href {\doibase 10.1038/nphys384} {\bibfield  {journal} {\bibinfo
  {journal} {Nature Physics}\ }\textbf {\bibinfo {volume} {2}},\ \bibinfo
  {pages} {620} (\bibinfo {year} {2006})}\BibitemShut {NoStop}%
\bibitem [{\citenamefont {Young}\ and\ \citenamefont {Kim}(2009)}]{Young2008}%
  \BibitemOpen
  \bibfield  {author} {\bibinfo {author} {\bibfnamefont {A.~F.}\ \bibnamefont
  {Young}}\ and\ \bibinfo {author} {\bibfnamefont {P.}~\bibnamefont {Kim}},\
  }\href {\doibase 10.1038/nphys1198} {\bibfield  {journal} {\bibinfo
  {journal} {Nature Physics}\ }\textbf {\bibinfo {volume} {5}},\ \bibinfo
  {pages} {222} (\bibinfo {year} {2009})}\BibitemShut {NoStop}%
\bibitem [{\citenamefont {Cheianov}\ and\ \citenamefont
  {Fal'ko}(2006)}]{Cheianov2006}%
  \BibitemOpen
  \bibfield  {author} {\bibinfo {author} {\bibfnamefont {V.~V.}\ \bibnamefont
  {Cheianov}}\ and\ \bibinfo {author} {\bibfnamefont {V.~I.}\ \bibnamefont
  {Fal'ko}},\ }\href {\doibase 10.1103/PhysRevB.74.041403} {\bibfield
  {journal} {\bibinfo  {journal} {Physical Review B}\ }\textbf {\bibinfo
  {volume} {74}},\ \bibinfo {pages} {041403} (\bibinfo {year}
  {2006})}\BibitemShut {NoStop}%
\bibitem [{\citenamefont {Sutar}\ \emph {et~al.}(2012)\citenamefont {Sutar},
  \citenamefont {Comfort}, \citenamefont {Liu}, \citenamefont {Taniguchi},
  \citenamefont {Watanabe},\ and\ \citenamefont {Lee}}]{Sutar2012}%
  \BibitemOpen
  \bibfield  {author} {\bibinfo {author} {\bibfnamefont {S.}~\bibnamefont
  {Sutar}}, \bibinfo {author} {\bibfnamefont {E.~S.}\ \bibnamefont {Comfort}},
  \bibinfo {author} {\bibfnamefont {J.}~\bibnamefont {Liu}}, \bibinfo {author}
  {\bibfnamefont {T.}~\bibnamefont {Taniguchi}}, \bibinfo {author}
  {\bibfnamefont {K.}~\bibnamefont {Watanabe}}, \ and\ \bibinfo {author}
  {\bibfnamefont {J.~U.}\ \bibnamefont {Lee}},\ }\href {\doibase
  10.1021/nl3011897} {\bibfield  {journal} {\bibinfo  {journal} {Nano Letters}\
  }\textbf {\bibinfo {volume} {12}},\ \bibinfo {pages} {4460} (\bibinfo {year}
  {2012})}\BibitemShut {NoStop}%
\bibitem [{\citenamefont {Sajjad}\ \emph {et~al.}(2012)\citenamefont {Sajjad},
  \citenamefont {Sutar}, \citenamefont {Lee},\ and\ \citenamefont
  {Ghosh}}]{Sajjad2012}%
  \BibitemOpen
  \bibfield  {author} {\bibinfo {author} {\bibfnamefont {R.~N.}\ \bibnamefont
  {Sajjad}}, \bibinfo {author} {\bibfnamefont {S.}~\bibnamefont {Sutar}},
  \bibinfo {author} {\bibfnamefont {J.~U.}\ \bibnamefont {Lee}}, \ and\
  \bibinfo {author} {\bibfnamefont {A.~W.}\ \bibnamefont {Ghosh}},\ }\href
  {\doibase 10.1103/PhysRevB.86.155412} {\bibfield  {journal} {\bibinfo
  {journal} {Phys. Rev. B}\ }\textbf {\bibinfo {volume} {86}},\ \bibinfo
  {pages} {155412} (\bibinfo {year} {2012})}\BibitemShut {NoStop}%
\bibitem [{\citenamefont {Abanin}\ and\ \citenamefont
  {Levitov}(2007)}]{Abanin2007}%
  \BibitemOpen
  \bibfield  {author} {\bibinfo {author} {\bibfnamefont {D.~A.}\ \bibnamefont
  {Abanin}}\ and\ \bibinfo {author} {\bibfnamefont {L.~S.}\ \bibnamefont
  {Levitov}},\ }\href {\doibase 10.1126/science.1144672} {\bibfield  {journal}
  {\bibinfo  {journal} {Science}\ }\textbf {\bibinfo {volume} {317}},\ \bibinfo
  {pages} {641} (\bibinfo {year} {2007})}\BibitemShut {NoStop}%
\bibitem [{\citenamefont {Williams}\ \emph {et~al.}(2007)\citenamefont
  {Williams}, \citenamefont {DiCarlo},\ and\ \citenamefont
  {Marcus}}]{Williams2007}%
  \BibitemOpen
  \bibfield  {author} {\bibinfo {author} {\bibfnamefont {J.~R.}\ \bibnamefont
  {Williams}}, \bibinfo {author} {\bibfnamefont {L.}~\bibnamefont {DiCarlo}}, \
  and\ \bibinfo {author} {\bibfnamefont {C.~M.}\ \bibnamefont {Marcus}},\
  }\href {\doibase 10.1126/science.1144657} {\bibfield  {journal} {\bibinfo
  {journal} {Science}\ }\textbf {\bibinfo {volume} {317}},\ \bibinfo {pages}
  {638} (\bibinfo {year} {2007})}\BibitemShut {NoStop}%
\bibitem [{\citenamefont {Klimov}\ \emph {et~al.}(2015)\citenamefont {Klimov},
  \citenamefont {Le}, \citenamefont {Yan}, \citenamefont {Agnihotri},
  \citenamefont {Comfort}, \citenamefont {Lee}, \citenamefont {Newell},\ and\
  \citenamefont {Richter}}]{NikolaiN.Klimov2015a}%
  \BibitemOpen
  \bibfield  {author} {\bibinfo {author} {\bibfnamefont {N.~N.}\ \bibnamefont
  {Klimov}}, \bibinfo {author} {\bibfnamefont {S.~T.}\ \bibnamefont {Le}},
  \bibinfo {author} {\bibfnamefont {J.}~\bibnamefont {Yan}}, \bibinfo {author}
  {\bibfnamefont {P.}~\bibnamefont {Agnihotri}}, \bibinfo {author}
  {\bibfnamefont {E.}~\bibnamefont {Comfort}}, \bibinfo {author} {\bibfnamefont
  {J.~U.}\ \bibnamefont {Lee}}, \bibinfo {author} {\bibfnamefont {D.~B.}\
  \bibnamefont {Newell}}, \ and\ \bibinfo {author} {\bibfnamefont {C.~A.}\
  \bibnamefont {Richter}},\ }\href {\doibase 10.1103/PhysRevB.92.241301}
  {\bibfield  {journal} {\bibinfo  {journal} {Physical Review B}\ }\textbf
  {\bibinfo {volume} {92}},\ \bibinfo {pages} {241301} (\bibinfo {year}
  {2015})}\BibitemShut {NoStop}%
\bibitem [{\citenamefont {Rickhaus}\ \emph {et~al.}(2013)\citenamefont
  {Rickhaus}, \citenamefont {Maurand}, \citenamefont {Liu}, \citenamefont
  {Weiss}, \citenamefont {Richter},\ and\ \citenamefont
  {Sch{\"{o}}nenberger}}]{Rickhaus2013}%
  \BibitemOpen
  \bibfield  {author} {\bibinfo {author} {\bibfnamefont {P.}~\bibnamefont
  {Rickhaus}}, \bibinfo {author} {\bibfnamefont {R.}~\bibnamefont {Maurand}},
  \bibinfo {author} {\bibfnamefont {M.-H.}\ \bibnamefont {Liu}}, \bibinfo
  {author} {\bibfnamefont {M.}~\bibnamefont {Weiss}}, \bibinfo {author}
  {\bibfnamefont {K.}~\bibnamefont {Richter}}, \ and\ \bibinfo {author}
  {\bibfnamefont {C.}~\bibnamefont {Sch{\"{o}}nenberger}},\ }\href {\doibase
  10.1038/ncomms3342} {\bibfield  {journal} {\bibinfo  {journal} {Nature
  Communications}\ }\textbf {\bibinfo {volume} {4}},\ \bibinfo {pages} {2342}
  (\bibinfo {year} {2013})}\BibitemShut {NoStop}%
\bibitem [{\citenamefont {Groth}\ \emph {et~al.}(2014)\citenamefont {Groth},
  \citenamefont {Wimmer}, \citenamefont {Akhmerov},\ and\ \citenamefont
  {Waintal}}]{Groth2014}%
  \BibitemOpen
  \bibfield  {author} {\bibinfo {author} {\bibfnamefont {C.~W.}\ \bibnamefont
  {Groth}}, \bibinfo {author} {\bibfnamefont {M.}~\bibnamefont {Wimmer}},
  \bibinfo {author} {\bibfnamefont {A.~R.}\ \bibnamefont {Akhmerov}}, \ and\
  \bibinfo {author} {\bibfnamefont {X.}~\bibnamefont {Waintal}},\ }\href
  {\doibase 10.1088/1367-2630/16/6/063065} {\bibfield  {journal} {\bibinfo
  {journal} {New Journal of Physics}\ }\textbf {\bibinfo {volume} {16}},\
  \bibinfo {pages} {063065} (\bibinfo {year} {2014})}\BibitemShut {NoStop}%
\bibitem [{\citenamefont {Low}\ and\ \citenamefont
  {Appenzeller}(2009)}]{Low2009b}%
  \BibitemOpen
  \bibfield  {author} {\bibinfo {author} {\bibfnamefont {T.}~\bibnamefont
  {Low}}\ and\ \bibinfo {author} {\bibfnamefont {J.}~\bibnamefont
  {Appenzeller}},\ }\href {\doibase 10.1103/PhysRevB.80.155406} {\bibfield
  {journal} {\bibinfo  {journal} {Physical Review B - Condensed Matter and
  Materials Physics}\ }\textbf {\bibinfo {volume} {80}},\ \bibinfo {pages} {1}
  (\bibinfo {year} {2009})}\BibitemShut {NoStop}%
\bibitem [{\citenamefont {LaGasse}\ and\ \citenamefont
  {Lee}(2016)}]{LaGasse2016}%
  \BibitemOpen
  \bibfield  {author} {\bibinfo {author} {\bibfnamefont {S.~W.}\ \bibnamefont
  {LaGasse}}\ and\ \bibinfo {author} {\bibfnamefont {J.~U.}\ \bibnamefont
  {Lee}},\ }\href {\doibase 10.1103/PhysRevB.94.165312} {\bibfield  {journal}
  {\bibinfo  {journal} {Phys. Rev. B}\ }\textbf {\bibinfo {volume} {94}},\
  \bibinfo {pages} {165312} (\bibinfo {year} {2016})}\BibitemShut {NoStop}%
\bibitem [{\citenamefont {B{\"{u}}ttiker}(1988)}]{Buttiker1988}%
  \BibitemOpen
  \bibfield  {author} {\bibinfo {author} {\bibfnamefont {M.}~\bibnamefont
  {B{\"{u}}ttiker}},\ }\href {\doibase 10.1103/PhysRevB.38.9375} {\bibfield
  {journal} {\bibinfo  {journal} {Physical Review B}\ }\textbf {\bibinfo
  {volume} {38}},\ \bibinfo {pages} {9375} (\bibinfo {year}
  {1988})}\BibitemShut {NoStop}%
\bibitem [{\citenamefont {Beenakker}(2008)}]{Beenakker2008a}%
  \BibitemOpen
  \bibfield  {author} {\bibinfo {author} {\bibfnamefont {C.~W.~J.}\
  \bibnamefont {Beenakker}},\ }\href {\doibase 10.1103/RevModPhys.80.1337}
  {\bibfield  {journal} {\bibinfo  {journal} {Reviews of Modern Physics}\
  }\textbf {\bibinfo {volume} {80}},\ \bibinfo {pages} {1337} (\bibinfo {year}
  {2008})}\BibitemShut {NoStop}%
\bibitem [{\citenamefont {Williams}\ and\ \citenamefont
  {Marcus}(2011)}]{Williams2011}%
  \BibitemOpen
  \bibfield  {author} {\bibinfo {author} {\bibfnamefont {J.~R.}\ \bibnamefont
  {Williams}}\ and\ \bibinfo {author} {\bibfnamefont {C.~M.}\ \bibnamefont
  {Marcus}},\ }\href {\doibase 10.1103/PhysRevLett.107.046602} {\bibfield
  {journal} {\bibinfo  {journal} {Physical Review Letters}\ }\textbf {\bibinfo
  {volume} {107}},\ \bibinfo {pages} {046602} (\bibinfo {year}
  {2011})}\BibitemShut {NoStop}%
\bibitem [{\citenamefont {Rickhaus}\ \emph {et~al.}(2015)\citenamefont
  {Rickhaus}, \citenamefont {Makk}, \citenamefont {Liu}, \citenamefont
  {T{\'{o}}v{\'{a}}ri}, \citenamefont {Weiss}, \citenamefont {Maurand},
  \citenamefont {Richter},\ and\ \citenamefont
  {Sch{\"{o}}nenberger}}]{Rickhaus2015b}%
  \BibitemOpen
  \bibfield  {author} {\bibinfo {author} {\bibfnamefont {P.}~\bibnamefont
  {Rickhaus}}, \bibinfo {author} {\bibfnamefont {P.}~\bibnamefont {Makk}},
  \bibinfo {author} {\bibfnamefont {M.-H.}\ \bibnamefont {Liu}}, \bibinfo
  {author} {\bibfnamefont {E.}~\bibnamefont {T{\'{o}}v{\'{a}}ri}}, \bibinfo
  {author} {\bibfnamefont {M.}~\bibnamefont {Weiss}}, \bibinfo {author}
  {\bibfnamefont {R.}~\bibnamefont {Maurand}}, \bibinfo {author} {\bibfnamefont
  {K.}~\bibnamefont {Richter}}, \ and\ \bibinfo {author} {\bibfnamefont
  {C.}~\bibnamefont {Sch{\"{o}}nenberger}},\ }\href {\doibase
  10.1038/ncomms7470} {\bibfield  {journal} {\bibinfo  {journal} {Nature
  Communications}\ }\textbf {\bibinfo {volume} {6}},\ \bibinfo {pages} {6470}
  (\bibinfo {year} {2015})}\BibitemShut {NoStop}%
\bibitem [{\citenamefont {Taychatanapat}\ \emph {et~al.}(2015)\citenamefont
  {Taychatanapat}, \citenamefont {Tan}, \citenamefont {Yeo}, \citenamefont
  {Watanabe}, \citenamefont {Taniguchi},\ and\ \citenamefont
  {{\"{O}}zyilmaz}}]{Taychatanapat2015}%
  \BibitemOpen
  \bibfield  {author} {\bibinfo {author} {\bibfnamefont {T.}~\bibnamefont
  {Taychatanapat}}, \bibinfo {author} {\bibfnamefont {J.~Y.}\ \bibnamefont
  {Tan}}, \bibinfo {author} {\bibfnamefont {Y.}~\bibnamefont {Yeo}}, \bibinfo
  {author} {\bibfnamefont {K.}~\bibnamefont {Watanabe}}, \bibinfo {author}
  {\bibfnamefont {T.}~\bibnamefont {Taniguchi}}, \ and\ \bibinfo {author}
  {\bibfnamefont {B.}~\bibnamefont {{\"{O}}zyilmaz}},\ }\href {\doibase
  10.1038/ncomms7093} {\bibfield  {journal} {\bibinfo  {journal} {Nature
  Communications}\ }\textbf {\bibinfo {volume} {6}},\ \bibinfo {pages} {6093}
  (\bibinfo {year} {2015})}\BibitemShut {NoStop}%
\bibitem [{\citenamefont {Zarenia}\ \emph {et~al.}(2013)\citenamefont
  {Zarenia}, \citenamefont {Pereira}, \citenamefont {Peeters},\ and\
  \citenamefont {Farias}}]{Zarenia2013}%
  \BibitemOpen
  \bibfield  {author} {\bibinfo {author} {\bibfnamefont {M.}~\bibnamefont
  {Zarenia}}, \bibinfo {author} {\bibfnamefont {J.~M.}\ \bibnamefont
  {Pereira}}, \bibinfo {author} {\bibfnamefont {F.~M.}\ \bibnamefont
  {Peeters}}, \ and\ \bibinfo {author} {\bibfnamefont {G.~A.}\ \bibnamefont
  {Farias}},\ }\href {\doibase 10.1103/PhysRevB.87.035426} {\bibfield
  {journal} {\bibinfo  {journal} {Physical Review B}\ }\textbf {\bibinfo
  {volume} {87}},\ \bibinfo {pages} {035426} (\bibinfo {year}
  {2013})}\BibitemShut {NoStop}%
\bibitem [{\citenamefont {Milovanovi{\'{c}}}\ \emph
  {et~al.}(2014{\natexlab{a}})\citenamefont {Milovanovi{\'{c}}}, \citenamefont
  {{Ramezani Masir}},\ and\ \citenamefont {Peeters}}]{Milovanovic2014a}%
  \BibitemOpen
  \bibfield  {author} {\bibinfo {author} {\bibfnamefont {S.~P.}\ \bibnamefont
  {Milovanovi{\'{c}}}}, \bibinfo {author} {\bibfnamefont {M.}~\bibnamefont
  {{Ramezani Masir}}}, \ and\ \bibinfo {author} {\bibfnamefont {F.~M.}\
  \bibnamefont {Peeters}},\ }\href {\doibase 10.1063/1.4896769} {\bibfield
  {journal} {\bibinfo  {journal} {Applied Physics Letters}\ }\textbf {\bibinfo
  {volume} {105}},\ \bibinfo {pages} {123507} (\bibinfo {year}
  {2014}{\natexlab{a}})}\BibitemShut {NoStop}%
\bibitem [{\citenamefont {{Kolasi{\'n}ski}}\ \emph {et~al.}(2016)\citenamefont
  {{Kolasi{\'n}ski}}, \citenamefont {{Mre{\'n}ca-Kolasi{\'n}ska}},\ and\
  \citenamefont {{Szafran}}}]{Kolasinski2016}%
  \BibitemOpen
  \bibfield  {author} {\bibinfo {author} {\bibfnamefont {K.}~\bibnamefont
  {{Kolasi{\'n}ski}}}, \bibinfo {author} {\bibfnamefont {A.}~\bibnamefont
  {{Mre{\'n}ca-Kolasi{\'n}ska}}}, \ and\ \bibinfo {author} {\bibfnamefont
  {B.}~\bibnamefont {{Szafran}}},\ }\href@noop {} {\bibfield  {journal}
  {\bibinfo  {journal} {ArXiv e-prints}\ } (\bibinfo {year} {2016})},\ \Eprint
  {http://arxiv.org/abs/1610.07566} {arXiv:1610.07566 [cond-mat.mes-hall]}
  \BibitemShut {NoStop}%
\bibitem [{\citenamefont {Carmier}\ \emph {et~al.}(2010)\citenamefont
  {Carmier}, \citenamefont {Lewenkopf},\ and\ \citenamefont
  {Ullmo}}]{Carmier2010}%
  \BibitemOpen
  \bibfield  {author} {\bibinfo {author} {\bibfnamefont {P.}~\bibnamefont
  {Carmier}}, \bibinfo {author} {\bibfnamefont {C.}~\bibnamefont {Lewenkopf}},
  \ and\ \bibinfo {author} {\bibfnamefont {D.}~\bibnamefont {Ullmo}},\ }\href
  {\doibase 10.1103/PhysRevB.81.241406} {\bibfield  {journal} {\bibinfo
  {journal} {Physical Review B}\ }\textbf {\bibinfo {volume} {81}},\ \bibinfo
  {pages} {241406} (\bibinfo {year} {2010})}\BibitemShut {NoStop}%
\bibitem [{\citenamefont {Carmier}\ \emph {et~al.}(2011)\citenamefont
  {Carmier}, \citenamefont {Lewenkopf},\ and\ \citenamefont
  {Ullmo}}]{Carmier2011}%
  \BibitemOpen
  \bibfield  {author} {\bibinfo {author} {\bibfnamefont {P.}~\bibnamefont
  {Carmier}}, \bibinfo {author} {\bibfnamefont {C.}~\bibnamefont {Lewenkopf}},
  \ and\ \bibinfo {author} {\bibfnamefont {D.}~\bibnamefont {Ullmo}},\ }\href
  {\doibase 10.1103/PhysRevB.84.195428} {\bibfield  {journal} {\bibinfo
  {journal} {Physical Review B}\ }\textbf {\bibinfo {volume} {84}},\ \bibinfo
  {pages} {195428} (\bibinfo {year} {2011})}\BibitemShut {NoStop}%
\bibitem [{\citenamefont {Patel}\ \emph {et~al.}(2012)\citenamefont {Patel},
  \citenamefont {Davies}, \citenamefont {Cheianov},\ and\ \citenamefont
  {Fal'ko}}]{Patel2012}%
  \BibitemOpen
  \bibfield  {author} {\bibinfo {author} {\bibfnamefont {A.~A.}\ \bibnamefont
  {Patel}}, \bibinfo {author} {\bibfnamefont {N.}~\bibnamefont {Davies}},
  \bibinfo {author} {\bibfnamefont {V.}~\bibnamefont {Cheianov}}, \ and\
  \bibinfo {author} {\bibfnamefont {V.~I.}\ \bibnamefont {Fal'ko}},\ }\href
  {\doibase 10.1103/PhysRevB.86.081413} {\bibfield  {journal} {\bibinfo
  {journal} {Phys. Rev. B}\ }\textbf {\bibinfo {volume} {86}},\ \bibinfo
  {pages} {081413} (\bibinfo {year} {2012})}\BibitemShut {NoStop}%
\bibitem [{\citenamefont {Davies}\ \emph {et~al.}(2012)\citenamefont {Davies},
  \citenamefont {Patel}, \citenamefont {Cortijo}, \citenamefont {Cheianov},
  \citenamefont {Guinea},\ and\ \citenamefont {Fal'ko}}]{Davies2012}%
  \BibitemOpen
  \bibfield  {author} {\bibinfo {author} {\bibfnamefont {N.}~\bibnamefont
  {Davies}}, \bibinfo {author} {\bibfnamefont {A.~A.}\ \bibnamefont {Patel}},
  \bibinfo {author} {\bibfnamefont {A.}~\bibnamefont {Cortijo}}, \bibinfo
  {author} {\bibfnamefont {V.}~\bibnamefont {Cheianov}}, \bibinfo {author}
  {\bibfnamefont {F.}~\bibnamefont {Guinea}}, \ and\ \bibinfo {author}
  {\bibfnamefont {V.~I.}\ \bibnamefont {Fal'ko}},\ }\href {\doibase
  10.1103/PhysRevB.85.155433} {\bibfield  {journal} {\bibinfo  {journal}
  {Physical Review B}\ }\textbf {\bibinfo {volume} {85}},\ \bibinfo {pages}
  {155433} (\bibinfo {year} {2012})}\BibitemShut {NoStop}%
\bibitem [{\citenamefont {Milovanović}\ \emph {et~al.}(2013)\citenamefont
  {Milovanović}, \citenamefont {{Ramezani Masir}},\ and\ \citenamefont
  {Peeters}}]{Milovanovic2013a}%
  \BibitemOpen
  \bibfield  {author} {\bibinfo {author} {\bibfnamefont {S.~P.}\ \bibnamefont
  {Milovanović}}, \bibinfo {author} {\bibfnamefont {M.}~\bibnamefont
  {{Ramezani Masir}}}, \ and\ \bibinfo {author} {\bibfnamefont {F.~M.}\
  \bibnamefont {Peeters}},\ }\href {\doibase 10.1063/1.4838557} {\bibfield
  {journal} {\bibinfo  {journal} {Applied Physics Letters}\ }\textbf {\bibinfo
  {volume} {103}},\ \bibinfo {pages} {233502} (\bibinfo {year}
  {2013})}\BibitemShut {NoStop}%
\bibitem [{\citenamefont {Milovanovi{\'{c}}}\ \emph
  {et~al.}(2014{\natexlab{b}})\citenamefont {Milovanovi{\'{c}}}, \citenamefont
  {{Ramezani Masir}},\ and\ \citenamefont {Peeters}}]{Milovanovic2014}%
  \BibitemOpen
  \bibfield  {author} {\bibinfo {author} {\bibfnamefont {S.~P.}\ \bibnamefont
  {Milovanovi{\'{c}}}}, \bibinfo {author} {\bibfnamefont {M.}~\bibnamefont
  {{Ramezani Masir}}}, \ and\ \bibinfo {author} {\bibfnamefont {F.~M.}\
  \bibnamefont {Peeters}},\ }\href {\doibase 10.1063/1.4863403} {\bibfield
  {journal} {\bibinfo  {journal} {Journal of Applied Physics}\ }\textbf
  {\bibinfo {volume} {115}},\ \bibinfo {pages} {043719} (\bibinfo {year}
  {2014}{\natexlab{b}})}\BibitemShut {NoStop}%
\bibitem [{\citenamefont {Shevtsov}\ \emph {et~al.}(2012)\citenamefont
  {Shevtsov}, \citenamefont {Carmier}, \citenamefont {Petitjean}, \citenamefont
  {Groth}, \citenamefont {Carpentier},\ and\ \citenamefont
  {Waintal}}]{Shevtsov2012}%
  \BibitemOpen
  \bibfield  {author} {\bibinfo {author} {\bibfnamefont {O.}~\bibnamefont
  {Shevtsov}}, \bibinfo {author} {\bibfnamefont {P.}~\bibnamefont {Carmier}},
  \bibinfo {author} {\bibfnamefont {C.}~\bibnamefont {Petitjean}}, \bibinfo
  {author} {\bibfnamefont {C.}~\bibnamefont {Groth}}, \bibinfo {author}
  {\bibfnamefont {D.}~\bibnamefont {Carpentier}}, \ and\ \bibinfo {author}
  {\bibfnamefont {X.}~\bibnamefont {Waintal}},\ }\href {\doibase
  10.1103/PhysRevX.2.031004} {\bibfield  {journal} {\bibinfo  {journal} {Phys.
  Rev. X}\ }\textbf {\bibinfo {volume} {2}},\ \bibinfo {pages} {031004}
  (\bibinfo {year} {2012})}\BibitemShut {NoStop}%
\bibitem [{\citenamefont {Liu}\ \emph {et~al.}(2015)\citenamefont {Liu},
  \citenamefont {Rickhaus}, \citenamefont {Makk}, \citenamefont
  {T{\'{o}}v{\'{a}}ri}, \citenamefont {Maurand}, \citenamefont {Tkatschenko},
  \citenamefont {Weiss}, \citenamefont {Sch{\"{o}}nenberger},\ and\
  \citenamefont {Richter}}]{Liu2015}%
  \BibitemOpen
  \bibfield  {author} {\bibinfo {author} {\bibfnamefont {M.-H.}\ \bibnamefont
  {Liu}}, \bibinfo {author} {\bibfnamefont {P.}~\bibnamefont {Rickhaus}},
  \bibinfo {author} {\bibfnamefont {P.}~\bibnamefont {Makk}}, \bibinfo {author}
  {\bibfnamefont {E.}~\bibnamefont {T{\'{o}}v{\'{a}}ri}}, \bibinfo {author}
  {\bibfnamefont {R.}~\bibnamefont {Maurand}}, \bibinfo {author} {\bibfnamefont
  {F.}~\bibnamefont {Tkatschenko}}, \bibinfo {author} {\bibfnamefont
  {M.}~\bibnamefont {Weiss}}, \bibinfo {author} {\bibfnamefont
  {C.}~\bibnamefont {Sch{\"{o}}nenberger}}, \ and\ \bibinfo {author}
  {\bibfnamefont {K.}~\bibnamefont {Richter}},\ }\href {\doibase
  10.1103/PhysRevLett.114.036601} {\bibfield  {journal} {\bibinfo  {journal}
  {Physical Review Letters}\ }\textbf {\bibinfo {volume} {114}},\ \bibinfo
  {pages} {036601} (\bibinfo {year} {2015})}\BibitemShut {NoStop}%
\bibitem [{\citenamefont {Reich}\ \emph {et~al.}(2002)\citenamefont {Reich},
  \citenamefont {Maultzsch}, \citenamefont {Thomsen},\ and\ \citenamefont
  {Ordej{\'{o}}n}}]{Reich2002}%
  \BibitemOpen
  \bibfield  {author} {\bibinfo {author} {\bibfnamefont {S.}~\bibnamefont
  {Reich}}, \bibinfo {author} {\bibfnamefont {J.}~\bibnamefont {Maultzsch}},
  \bibinfo {author} {\bibfnamefont {C.}~\bibnamefont {Thomsen}}, \ and\
  \bibinfo {author} {\bibfnamefont {P.}~\bibnamefont {Ordej{\'{o}}n}},\ }\href
  {\doibase 10.1103/PhysRevB.66.035412} {\bibfield  {journal} {\bibinfo
  {journal} {Physical Review B}\ }\textbf {\bibinfo {volume} {66}},\ \bibinfo
  {pages} {035412} (\bibinfo {year} {2002})}\BibitemShut {NoStop}%
\bibitem [{\citenamefont {Stegmann}\ \emph {et~al.}(2013)\citenamefont
  {Stegmann}, \citenamefont {Wolf},\ and\ \citenamefont
  {Lorke}}]{Stegmann2013}%
  \BibitemOpen
  \bibfield  {author} {\bibinfo {author} {\bibfnamefont {T.}~\bibnamefont
  {Stegmann}}, \bibinfo {author} {\bibfnamefont {D.~E.}\ \bibnamefont {Wolf}},
  \ and\ \bibinfo {author} {\bibfnamefont {A.}~\bibnamefont {Lorke}},\ }\href
  {\doibase 10.1088/1367-2630/15/11/113047} {\bibfield  {journal} {\bibinfo
  {journal} {New Journal of Physics}\ }\textbf {\bibinfo {volume} {15}},\
  \bibinfo {pages} {113047} (\bibinfo {year} {2013})}\BibitemShut {NoStop}%
\bibitem [{\citenamefont {Stegmann}\ and\ \citenamefont
  {Lorke}(2015)}]{Stegmann2015}%
  \BibitemOpen
  \bibfield  {author} {\bibinfo {author} {\bibfnamefont {T.}~\bibnamefont
  {Stegmann}}\ and\ \bibinfo {author} {\bibfnamefont {A.}~\bibnamefont
  {Lorke}},\ }\href {\doibase 10.1002/andp.201500124} {\bibfield  {journal}
  {\bibinfo  {journal} {Annalen der Physik}\ }\textbf {\bibinfo {volume}
  {527}},\ \bibinfo {pages} {723} (\bibinfo {year} {2015})}\BibitemShut
  {NoStop}%
\bibitem [{\citenamefont {Datta}(1995)}]{Datta1995}%
  \BibitemOpen
  \bibfield  {author} {\bibinfo {author} {\bibfnamefont {S.}~\bibnamefont
  {Datta}},\ }\href {\doibase 10.1017/CBO9780511805776} {\emph {\bibinfo
  {title} {{Electronic Transport in Mesoscopic Systems}}}}\ (\bibinfo
  {publisher} {Cambridge University Press},\ \bibinfo {address} {Cambridge},\
  \bibinfo {year} {1995})\ p.\ \bibinfo {pages} {393}\BibitemShut {NoStop}%
\end{thebibliography}
\end{document}